# ANAPOLE MODE EXCITATION IN NOVEL PERFORATED ALL-DIELECTRIC METAMATERIALS

*Anar K. Ospanova[1,2,3], Ivan V. Stenishchev[1,2,3], Alexey A. Basharin[1,2,3]*

[1] National University of Science and Technology (MISiS), Department of Theoretical Physics and Quantum Technologies, 119049 Moscow, Russia
[2] National University of Science and Technology (MISiS), The Laboratory of Superconducting metamaterials, 119049 Moscow, Russia
[3] Politecnico di Torino, Department of Electronic and Telecommunications, Torino 10129, Italy
*Corresponding author: alexey.basharin@misis.ru

*Dated 29 January 2018*

## ABSTRACT

In this paper, we concern on nonradiating properties of novel silicon metamaterials due to multipolar interaction. Destructive interference of electric dipole moment and the next term of multipolar decomposition - toroidal dipole moment, leads to nonradiating anapole mode, which is nowadays widely studying among electrodynamic phenomena. Therefore, we propose technically simple metamaterial design, which does not demand multistep fabrication. This kind of metamaterial is the promising in nanophotonics and plasmonics due to subtle sensing, nonradiative data transfer, Aharonov-Bohm effect and other demonstrations.

## INTRODUCTION

Metamaterials are composite media consisting of arranged subwavelength building blocks called metamolecules. These artificial structures are popular for their unnatural properties such as negative refractive index, strong field localization, cloaking, strong magnetic response, superlensing effect and others. Metamaterials are of high technological demand since their electromagnetic properties can be tuned easily by changing geometrical sizes and shapes [1-4]. Although metallic metamaterials are promising candidates for light manipulation, they inherently possess high dissipative losses. Since nanoscale devices gained popularity, there is another obstacle for metallic metamaterials in their size scaling limit in optical range [5, 6].

On the other hand, all-dielectric metamaterials are widely used due to several advantages over metallic metamaterials. The main difference between metallic and all-dielectric metamaterials is in the nature of induced currents. An illuminated electromagnetic wave induces displacement currents in all-dielectric metaparticles, which possess low dissipative losses. Additionally, all-dielectric metamaterials are scalable in optical frequency range and their resonances can be controlled by means of tailoring the permittivity and sizes of metaparticles. All-dielectric metamolecules provide strong magnetic response similar to those of plasmonic Split-Ring Resonators (SRR), but features less dissipative losses in visible and near IR range. Due to different resonance order of dielectric materials, all-dielectric metaparticles firstly exhibit magnetic dipole resonance in visible spectral range, and makes it possible to create "magnetic" light [7-11].

Among other dielectric materials, silicon is extensively investigated due to number of reasons. Specially arranged silicon metasurfaces and metamaterials are used for phase retardation, overcoming chromatic aberration for multiwavelength devices, Huygen's surfaces, beam shaping, biosensing and other. Additionally to their optical properties, silicon nanostructures are mechanically, electrically, optically and thermal robust [12-16].

Comprehensive experimental investigation of silicon nanoparticles in visible spectral range revealed the possibility of anapole mode establishment in such metamaterials. The term static anapole firstly introduced by Y. B. Zel'dovich in nuclear physics for description of weak interactions in nucleus [17]. In electrodynamics, dynamic anapole mode is destructive interference of electric and toroidal multipoles, resulting in extermination of far-field radiation. Here, toroidal multipole is the next kind of multipolar decomposition of currents and represents toroidal configuration of poloidal currents with corresponding closed head-to-tail magnetic loop. Toroidal response of this configuration oscillates in perpendicular to the magnetic loop surface [18-20]. The engineering challenge of anapole mode sustaining metamolecule is on realization of toroidal geometry that support dynamically induced and spatially confined magnetic loop. Especially, this problem is acute in optical frequencies as well as engineering of any other 3D structures in general. On the other hand, anapole mode is in the spotlight due to a number of features. Primarily, because of the ability to concentrate strong electromagnetic fields inside a point nonradiating source or scatterer and ensuring the suppression of radiation in an external field. Therefore, metamaterials with anapole mode demonstrated extremely high Q factor so that manifested themselves as perfect resonator [21-30].

All these properties lead to extensive research of all-dielectric and, particularly, silicon metamaterials and their many applications. 2D metamaterials or metasurfaces stand out due to the simplicity of fabrication and exhibits desirable optical properties [11, 22, 31-39]. Nevertheless, we should mention the Ref. [40], in which the possibility of a toroidal excitation in plasmonic toroidal metamaterials of the optical range has been shown. Commonly exploited technique is electron-beam lithography with subsequent reactive ion etching (RIE) or carving procedure or femtosecond laser [41]. Although most experimental works are based on this technique, their use is limited for spherical and ellipsoidal shaped nanodisks [42, 43]. Alternative techniques are laser printing of nanospheres and chemical vapor deposition. Both techniques are very sophisticated since nanosphere fabrication demands high precision and multi-step realization.

In this paper, we demonstrate novel silicon metamaterials sustaining anapole mode in visible spectral range. Therefore, this type metamaterial is promising since it does not require high-tech implementation.

## THE STRUCTURE OF THE SYSTEM

Our metamaterial consists of arranged metamolecules with clusters of four throughly perforated holes in silicon slab. The edges of surface have the similar length 200 nm and thickness of metamolecule is 100 nm. The diameter of each hole is 45 nm, period of clusters is 200 nm and center-to-center distance between holes is 55 nm. We assume linearly polarized incident wave normally directed onto top of metamaterial.

As of any other dielectric material, electromagnetic response of our silicon metamaterial is defined due to Mie resonances of displacement currents induced within cluster [44]. As can be seen from Fig. 1, metamolecule is illuminated from the top facet and both electric and magnetic counterparts of impinging wave propagates perpendicular to holes' axis. Tailored geometrical sizes and dielectric permittivity of metamaterial together with polarization and angle of incident wave leads to peculiar (desirable) form of displacement currents. In turn, these currents cause resonant electromagnetic scattering also called Mie resonance. This Mie resonance explains accurately electromagnetic properties of dielectric nanoparticles, especially the appearance of strong magnetic modes. Mie resonance also provides essential for toroidal response form of current distribution. Displacement currents **j** take the form of meridians on the toroidal surface

(like poloidal currents) and generates magnetic mode confined within the "imaginary" torus. This configuration creates electromagnetic analogue of toroidal moment in all-dielectric metamaterials. According to Fig. 1, the front of incident wave propagates along the holes' axis and its electric component **E** oscillates along holes. Consequently, displacement currents **j** circle holes and creates two loops of current: counterclockwise and clockwise. These current loops create circulated magnetic mode **m** restricted within these loops. This configuration expectedly creates toroidal response **T** oscillating in the direction of electric component of incident wave. The antiphase oscillation of toroidal **T** and electric **P** modes configuration in the relation of **P**=$ik$**T** creates anapole mode at resonant frequency.

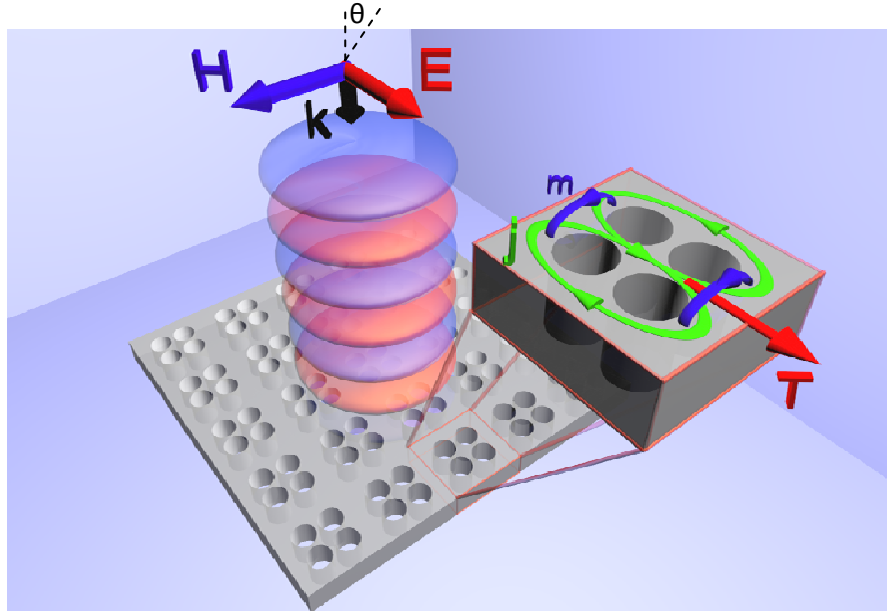

Fig. 1. Illustration of the proposed optical metamaterial supplemented with current and mode distribution within each metamolecule. Inset shows toroidal mode excitation; here **m** stands for magnetic dipole moment, **j** – electric current loops, **T** – toroidal dipole moment.

RESULTS

The electromagnetic properties of silicon slab is calculated by commercial Maxwell's equation solver HFSS using standard modeling approach, where the properties of whole structure represented by the parameters of their unit cell with correctly applied boundary conditions. Figure 2 depicts transmission spectrum of perforated along its thickness silicon slab near their resonant frequency. One can see full transmission peak at f=566.5 THz with amplitude |T|=1. Fig 3 provides field maps of absolute value and cutplane in the direction of wave propagation of both electric and magnetic component at resonant frequency f=566.5 THz. Thus, the electric field strongly concentrates between pairs of holes and preconditions formation of displacement current loops **j** mimicking poloidal currents of the gedanken torus (Fig. 3(a)). These currents create magnetic fields circulated above and below holes (Fig. 3(b)). Moreover, magnetic field strongly concentrates on the top and bottom facets of metamolecule and rotates clockwise (Fig. 3(b)). Accordingly, we expect that such electromagnetic configuration of displacement currents and fields induce toroidal response **T** within metamolecule oscillating up and down on the direction of electric component **E**.

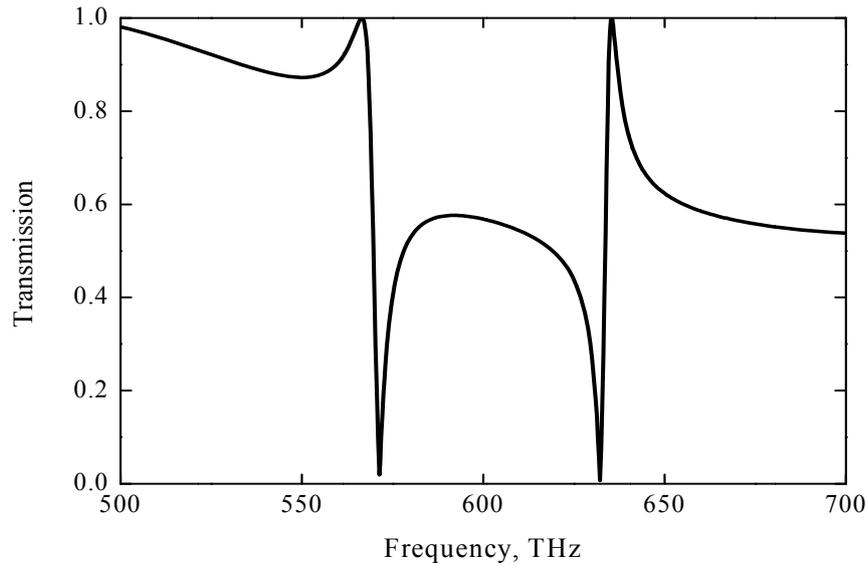

Fig 2. Transmission spectrum for metamolecule depicted on Fig. 1. Sharp transparency peak corresponds to f=566.5 THz.

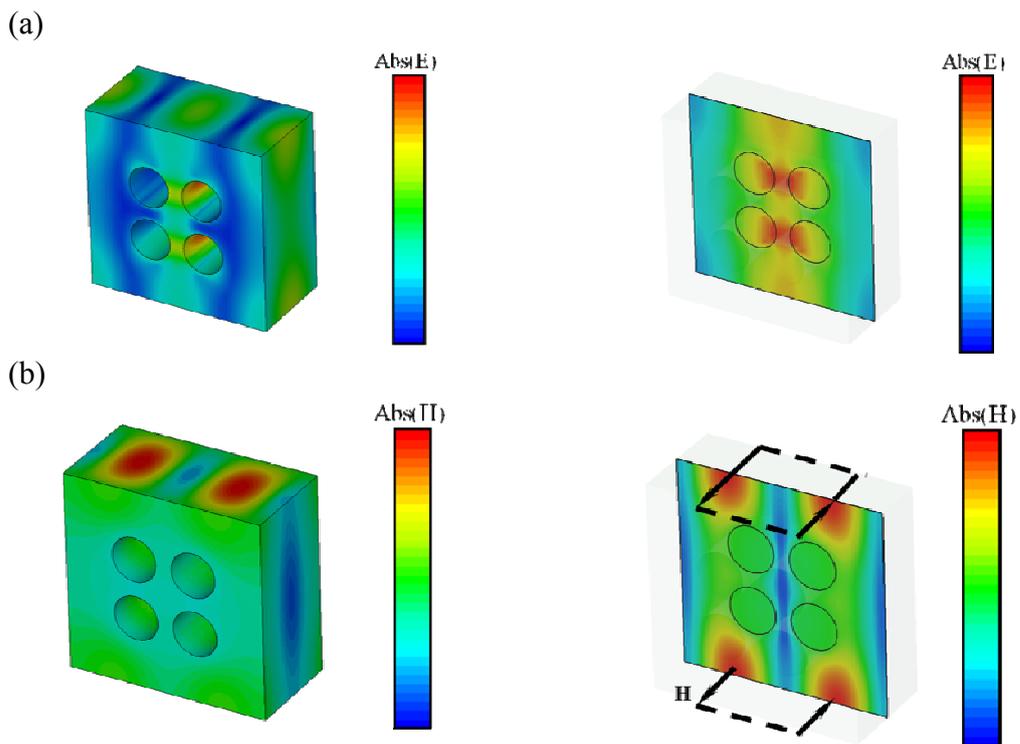

Fig. 3 Field maps of (a) absolute value of electric field and in cross-section of silicon perforated slab and (b) absolute value of magnetic field intensities field and in cross-section of silicon perforated slab.

The origin of electric and magnetic fields on the resonance frequency is defined by Mie-type modes exciting in dielectric metamolecules due to displacement currents and such field configuration corresponds to anapole mode excitation, which features strong electric field localization between holes and magnetic fields with low radiation into far field that intrinsic for anapole mode excitation owing to destructive interference of electric and toroidal moments.

To confirm our assumption, we carry out multipolar decomposition of displacement currenst up to second order multipoles: electric **P** and magnetic dipoles **M**, toroidal dipole **T**, electric **Qe** and magnetic **Qm** quadrupoles illustrated in Fig 4. Thus, close to resonant frequency there is compelling electric dipole **P** and toroidal dipole **T** which at resonant frequency f=566.5 THz come into oscillation of **P** = $ik$**T** so that establish anapole mode. Additionally, we can see rather high value of magnetic dipole moments throughout a given range which is below to 10 times. Such strong magnetic dipolar response contributes on widening of transmission peak so that suppress of Q factor due to radiative nature of magnetic mode. Electric and magnetic quadrupole moments are much lower than electric and toroidal dipole moments, so they do not induce considerable contribution.

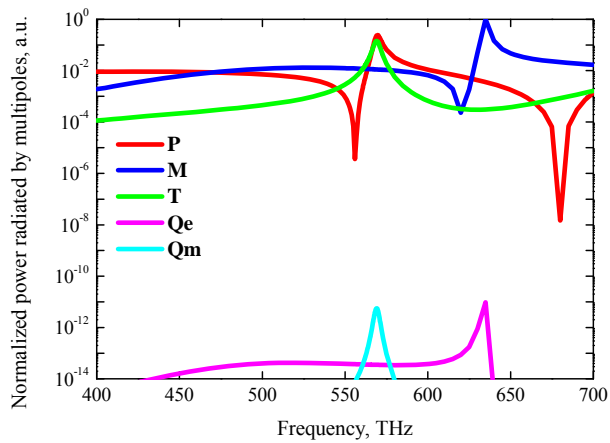

Fig. 4. Normalized power radiated (scattered) by metamaterial of near-field distribution of metamolecule up to second order multipoles.

Furthermore, we consider different depth of holes and inquire at which value anapole mode disappears. In figure 5, we provide transmission spectrum of metamolecule with holes of different depth h for cases of h = 30 nm, 40 nm, 50 nm, 80 nm and 100 nm. One can see that shortening the depth of holes leads to redshift of resonant frequency. It also accompanied by widening of transmission peak for low value of depth h. Figure 6 illustrates multipolar decomposition of near field for all presented cases of depth h and anapole mode disappears for h=30 nm. Multipolar decomposition for throughly perforated metamaterial presented in Fig. 4.

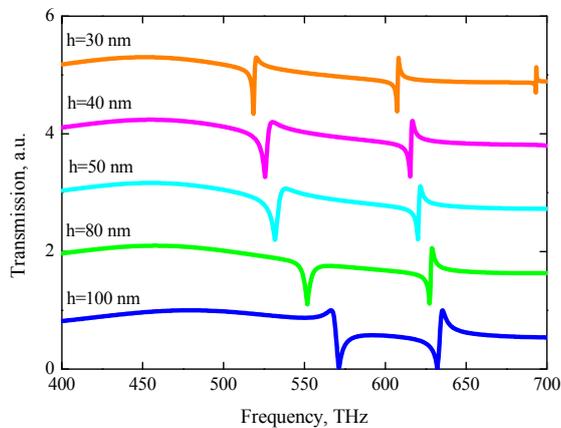

Fig 5. Transmission versus frequency for various depths h of silicon slab at resonant frequency.

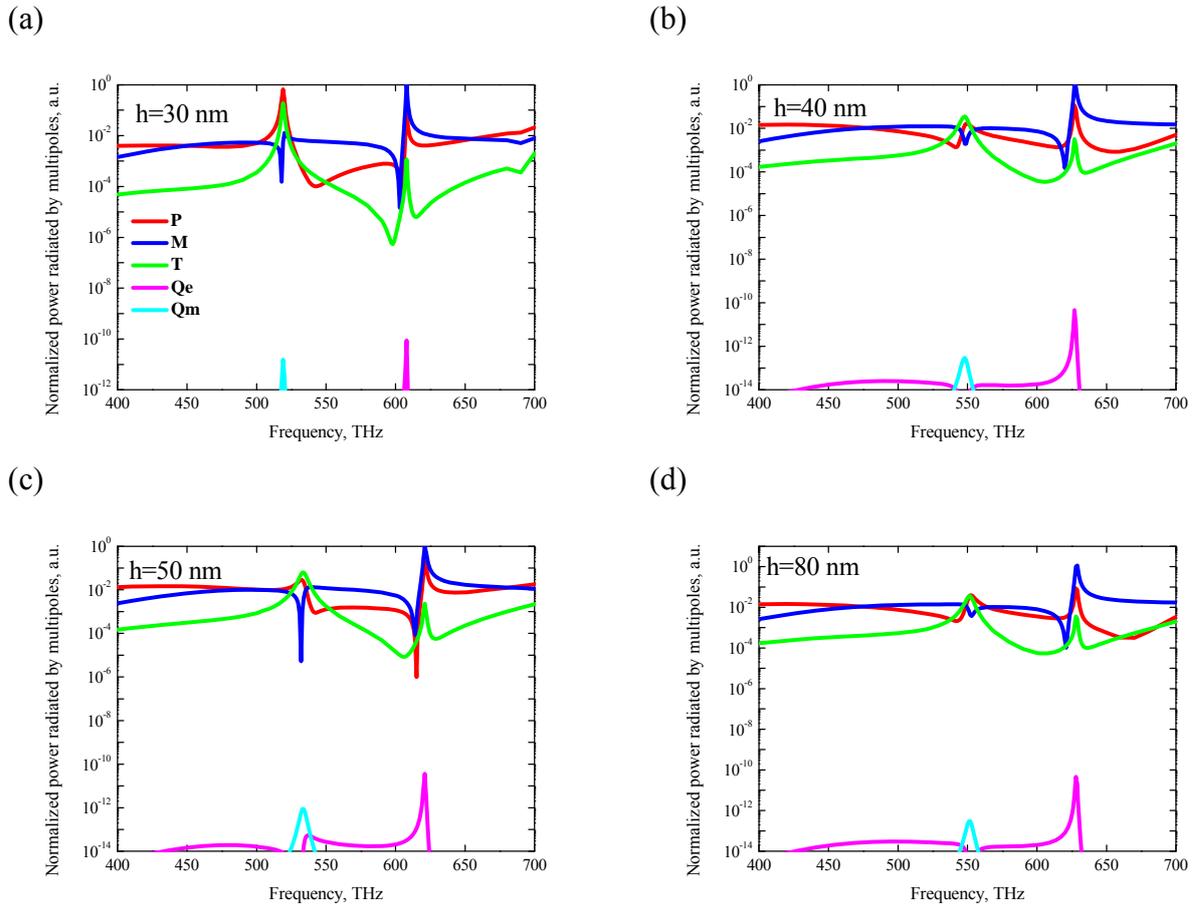

Fig 6. Contributions of the six strongest multipolar excitations to the normalized power radiated (scattered) by metamaterial for various depths h=30 nm, 40 nm, 50 nm, 80 nm of silicon slab at resonant frequency.

In addition, we estimated evaluation of transmission spectra for different radius d and incident angle θ. In figure 7 (a), we present transmission spectra for diameters d=30 nm, 45 nm, 50 nm, 60 nm, 70 nm and one can see that larger diameters lead to blueshift of transmission peak. Higher diameters d=50 nm, 60 nm, 70 nm lead to widening of transmission maximum, i.e. decrease of Q-factor. The case of small diameter (d=30 nm) corresponds to blueshift of resonance frequency. Although d=30 nm shows the sharpest transmission peak, this case obstacles difficulties in fabrication. Fig 7 (b) enables to compare how the angle of incidence affects transmission spectrum. We consider the case of incidence angles θ= $0^0$, $20^0$, $40^0$, $60^0$, $80^0$. The graph shows that gradually increase of angle of incidence leads to redshift of transmission peak and narrowing its resonance.

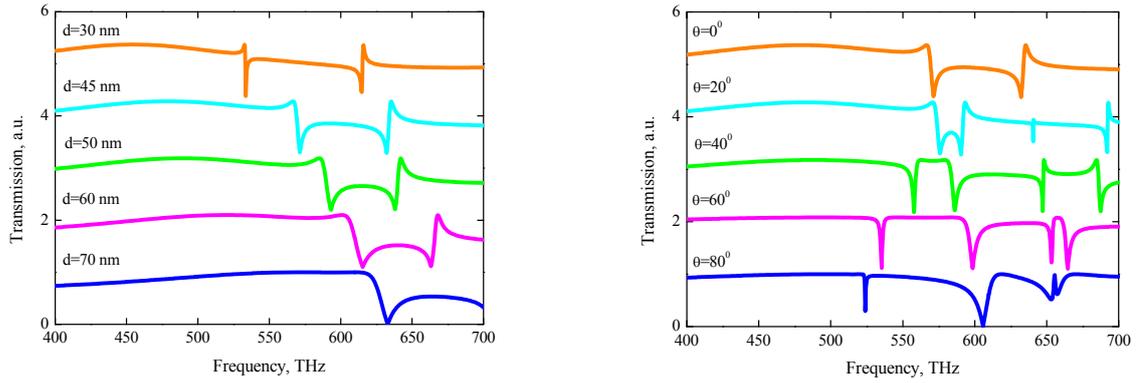

Fig 7. Transmission versus frequency for various diameters d of holes (a) and angles θ of incidence.

DISCUSSION

In this work, we propose novel type of anapole mode sustaining metamaterial in optics. Such metamaterials awaited to exhibit fascinating effects in visible spectral range. Papers of recent years reports more and more effects on anapole mode establishment in metamaterials. They have a great potential for application as near-field cloaked sources/sensors so that embodies ambitions of invisible movement. Such metamaterials makes foundation for dynamic Aharonov-Bohm demonstration in visible spectral range, which makes secure communications one-step closer.

The main purpose of our work is how all-dielectric anapole metamaterial can be realized due to simpler way. Our metamaterial apparently advantages in fabrication and means sample fabrication consisting of holes perforation, for example, FIB (Focused Ion Beam) method, which is applicable in nanoscale. Silicon is the most preferable material in optics and useful for well-pronounced effects demonstration. Besides, silicon is reasonably cheap and easy to obtain. Since silicon particles are electrically and thermal robust, it is convenient for nonlinear optical effects.

Since FIB fabrication takes place in one step, this kind metamaterial benefits both in time and fabrication price. On the other hand, our structure is promising for application in optical bio/cloaked sensing. Any liquid poured inside such cluster takes form of holes and shifts spectral response of metamaterial. This simplifies specification of liquid nature so that advances biosensors. On the other hand, this metamaterial can be used as a part of anapole cloaking due to field localization within a slab. Integration of metamaterial concept with planar waveguide theory is awaited to extend waveguide application in the field of cloaking. Our approach can be adapted for cloaking waveguides. Indeed, perforated silicon waveguide can be considered as transparent system. Accordingly, side wave obliged waveguide transparent through the waveguide due to anapole mode excitation. Therefore, such anapole mode sustaining metamolecule is useful for future of nonradiating sources and subwavelength resonators with extremely high Q factor.

CONCLUSION

In this paper, we demonstrated for the first time anapole mode sustaining all-dielectric metamaterial in visible spectral range. Furthermore, this type metamaterial advantages in fabrication, i.e. in time and price among other known nanoscale metamaterials. Our metamaterial is made by perforating holes in solid silicon slab. Besides, our metamaterial features considerably high Q and strong field localization which makes it outstanding candidate for perfect resonator. Field distribution at resonant frequency showed inherent to anapole mode strong field localization within metamolecule. In confirmation of our assumption, we carried out multipolar decomposition of near field by strongest multipoles. Our results proved that transparency effect at given resonant frequency, indeed, underpinned by anapole mode excitation. Therefore, we provided transmission spectra for various diameters of holes and angles of incidence and observed their evolution on shift of transmission peak. Our metamaterial paves the way for advanced optical devices on the base of all-dielectric metamaterials. Besides inherent low dissipative losses and strong anapole response, such optical metamaterial supposed to demonstrate subtle sensing, nonradiative data transfer, Aharonov-Bohm effect and other tempting applications in nanophotonics.